\documentclass[a4paper,11pt]{article}
\pdfoutput=1 
\usepackage{jheppub}
\usepackage[T1]{fontenc} 
\usepackage{color}
\usepackage{fancybox}
\usepackage{amsmath}
\usepackage{amsfonts}
\usepackage{slashed}
\usepackage{amssymb}
\usepackage{indentfirst}
\usepackage{stmaryrd}
\usepackage{subcaption}
\usepackage{graphicx}
\usepackage{cleveref}
\usepackage{textcomp}
\usepackage[toc]{appendix}
\usepackage[font={small,it}]{caption}
\usepackage[justification=centering]{caption}

\allowdisplaybreaks

\newcommand{\be}{\begin{equation}}
\newcommand{\bea}{\begin{eqnarray}}

\newcommand{\ee}{\end{equation}}
\newcommand{\eea}{\end{eqnarray}}

\title{\boldmath Notes on the $D=11$ pure spinor superparticle}

\author{Nathan Berkovits$^{a}$,  Eduardo Casali$^{b,d}$, Max Guillen$^{a,b,c}$, Lionel Mason$^{b}$}

\affiliation{$^{a}$ICTP South American Institute for Fundamental Reserch\\
Instituto de F\'{i}sica Te\'{o}rica, UNESP-Universidade Estadual Paulista\\ R. Dr. Bento T. Ferraz 271, Bl. II, S\~{a}o Paulo 01140-070, SP, Brazil\\
$^{b}$ Mathematical Institute,
University of Oxford\\
Andrew Wiles Building, Woodstock Road, Oxford OX2 6GG, United Kingdom.\\
$^{c}$ Perimeter Institute for Theoretical Physics\\
31 Caroline St N Waterloo, Ontario N2L 2Y5, Canada\\
$^{d}$ Center for Quantum Mathematics and Physics\\
Department of Physics, University of California, Davis, CA 95616 USA
}

\emailAdd{nathan.berkovits@unesp.br,ecasali@ucdavis.edu,luis.max@unesp.br,\\lmason@maths.ox.ac.uk}

\abstract{The $D=11$ pure spinor superparticle has been shown to describe linearized $D=11$ supergravity in a manifestly covariant way. A number of authors have proposed that its correlation functions be used to compute  amplitudes. The use of the scalar structure of the eleven-dimensional pure spinor top cohomology introduces a natural measure for computing such correlation functions. This prescription requires the construction of ghost number one and zero vertex operators. In these notes, we construct explicitly a ghost number one vertex operator but show the incompatibiliy of a ghost number zero vertex operator satisfying a standard descent equation for $D=11$ supergravity.}

\keywords{Supergravity, Superparticle, Pure spinors.}

\begin{document} 
\hfill{}
\maketitle

\section{Introduction}
Since eleven-dimensional supergravity \cite{Cremmer:1978km} is the low-energy limit of M-theory, worldline methods for computing $D=11$ supergravity amplitudes may lead to new insights into M-theory.  Worldline methods using the $D=11$ superparticle in light-cone gauge were developed in \cite{Green:1999by}, but a super-Poincar\'e covariant description of the
$D=11$ superparticle using pure spinors could be more powerful for making cancellations manifest and simplifying amplitude computations. 

Pure spinors were introduced in $D=10$ and $D=11$ supersymmetric field theories in \cite{HOWE1991141} and \cite{Howe:1991bx}, and were introduced in the context of superstring theory in \cite{Berkovits:2000fe} as extra dynamical variables on the worldsheet. These extra variables allowed super-Poincar\'e covariant quantization using a simple BRST operator and simplified the computation of multiloop scattering amplitudes as compared to the other superstring formalisms.
Pure spinors have also been  used for worldline field theory computations in quantum field theories \cite{Bjornsson:2010wm,Bjornsson:2010wu} where the ultraviolet behavior of the 4-point amplitude for ten-dimensional super Yang-Mills and Type II supergravity up to 5-loops was studied using power counting arguments. 


The eleven-dimensional analogs of pure spinors that are discussed here for the superparticle were introduced in \cite{Berkovits:2002uc} and used by \cite{Anguelova:2004pg} to set up a framework for computing $N$-point correlation functions at tree and loop level using a worldline field theory framework. Some higher-loop computations using the non-minimal $D=11$ pure spinor formalism of \cite{Cederwall:2010tn} have been performed in \cite{Cederwall:2012es,Karlsson:2014xva}. 

In this paper we provide evidence that contradicts some of the assumptions made in \cite{Anguelova:2004pg}. We construct the ghost number one vertex operator as a perturbation of the BRST operator.  This will be BRST invariant only when the $D=11$ supergravity equations of motion are imposed. This vertex operator takes the form
\begin{eqnarray}
U^{(1)} &=& \lambda^{\alpha}[h_{\alpha}{}^{a}P_{a} - h_{\alpha}{}^{\beta}d_{\beta} + \Omega_{\alpha ab}N^{ab}]
\end{eqnarray}
where $h_{\alpha}{}^{a}$, $h_{\alpha}{}^{\beta}$, $\Omega_{\alpha ab}$ come from small perturbations of the eleven-dimensional vielbeins and the structure equations of linearized $D=11$ supergravity. They satisfy equations of motion and gauge freedoms arising from the $D=11$ supergravity dynamical constraints.  These  determine their full $\theta$-expansions as explained in \cite{Tsimpis:2004gq} and these are required  in correlation function prescriptions involving $U^{(1)}$.

The eleven-dimensional pure spinor prescription for computing 
tree-level $N$-point correlation functions given in \cite{Anguelova:2004pg} requires the existence of a ghost number zero vertex operator satisfying the standard descent relation
\begin{eqnarray}\label{int1}
\{Q, V^{(0)}\} &=& [H, U^{(1)}]
\end{eqnarray}
where $H = P^{2}$ is the particle Hamiltonian. In this paper we will show that eqn. \eqref{int1} is incompatible with linearized $D=11$ supergravity and discuss some possible ways to fix this problem without going into details.  Our investigations were motivated by an attempt to extend the 11-dimensional pure-spinor superparticle to an 11-dimensional pure-spinor ambitwistor-string following \cite{Mason:2013sva,Berkovits:2013xba}, but the  incompatibility of \eqref{int1} appears to be an obstruction.

\vspace{2mm}
The paper is organized as follows. In section 2 we review the $D=11$ pure spinor superparticle. In section 3 we construct the ghost number one vertex operator by requiring that the pure spinor BRST operator be nilpotent at first order as an on-shell geometric deformation of the BRST charge. In section 4, we show the inconsistency between the descent equation \eqref{int1} relating ghost number one and zero vertex operators and the structure equations of $D=11$ supergravity. Finally, we give a self-contained review of the superspace formulation of $D=11$ supergravity in Appendix \ref{apA}.

\section{$D=11$ pure spinor superparticle}
The eleven-dimensional pure spinor superparticle action in a flat background is given by \cite{Berkovits:2002uc,Guillen:2017mte}
\begin{equation}\label{eeq1}
S = \int d\tau [P_{m}\partial_{\tau}X^{m} + p_{\mu}\partial_{\tau}\theta^{\mu} + w_{\alpha}\partial_{\tau}\lambda^{\alpha} - \frac{1}{2}P^{m}P_{m}]\, .
\end{equation}
We will use lowercase letters from the beginning/middle of the Greek alphabet to denote $SO(10,1)$ tangent/curved-space spinor indices and we will let lowercase letters from the beginning/midle of the Latin alphabet denote $SO(10,1)$ tangent/curved-space vector indices. The superspace fermionic coordinate $\theta^{\mu}$ is an $SO(10,1)$ Majorana spinor and $p_{\mu}$ is its respective canonical conjugate momentum, and $P_{m}$ is the momentum for $X^m$. The variable $\lambda^{\alpha}$ is a $D=11$ pure spinor variable\footnote{Note that we do not require $\lambda\Gamma^{ab}\lambda=0$ which would also be imposed by Cartan's definition of purity. } satisfying $\lambda\Gamma^{a}\lambda = 0$, and $w_{\alpha}$ is its conjugate momentum which is defined up to the gauge transformation $\delta w_{\alpha} = (\Gamma^{a}\lambda)_{\alpha}r_{m}$, for an arbitrary gauge parameter $r_{m}$. The $SO(10,1)$ gamma matrices denoted by $\Gamma^{a}$ satisfy the Clifford algebra $(\Gamma^{a})_{\alpha\beta}(\Gamma^{b})^{\beta\sigma} + (\Gamma^{b})_{\alpha\beta}(\Gamma^{a})^{\beta\sigma} = 2\eta^{ab}\delta^{\sigma}_{\alpha}$. In $D=11$ dimensions there exist an antisymmetric spinor metric $C_{\alpha\beta}$ (and its inverse $(C^{-1})^{\alpha\beta}$) which allows us to lower (and raise) spinor indices. 

The BRST operator associated to this theory is defined to be
\begin{eqnarray}
Q &=& \lambda^{\alpha}d_{\alpha}
\end{eqnarray}
where $d_{\alpha} = p_{\alpha} - \frac{1}{2}(\Gamma^{a}\theta)_{\alpha}P_{a}$ are the fermionic constraints of the $D=11$ Brink-Schwarz-like superparticle. The nilpotency of this operator follows immediately from the pureness of $\lambda^{\alpha}$, and thus physical states can be defined as elements of its cohomology. As shown in \cite{Berkovits:2002uc}, this BRST cohomology turns out to describe linearized $D=11$ supergravity in its Batalin-Vilkovisky formulation. The $D=11$ supergravity physical fields are found in the ghost number three sector of the cohomology. To see this, one can write the most general ghost number three superfield
\begin{eqnarray}
U^{(3)} &=& \lambda^{\alpha}\lambda^{\beta}\lambda^{\delta}C_{\alpha\beta\delta}(x,\theta)
\end{eqnarray}
The physical state conditions will constrain the functional form of $C_{\alpha\beta\delta}$ to be
\begin{eqnarray}
C_{\alpha\beta\delta} &=& (\gamma^{a}\theta)_{\alpha}(\gamma^{b}\theta)_{\beta}(\gamma^{c}\theta)_{\delta}C_{abc}(x) + (\gamma^{(a}\theta)_{\alpha}(\gamma^{b)c}\theta)_{\beta}(\gamma_{c}\theta)_{\delta}h_{ab}(x)\nonumber\\
&& + (\gamma^{b}\theta)_{\alpha}[(\gamma^{c}\theta)_{\beta}(\gamma^{d}\theta)_{\delta}(\theta\gamma_{cd})_{\epsilon} - (\gamma_{cd}\theta)_{\beta}(\gamma_{c}\theta)_{\delta}(\gamma_{d}\theta)_{\epsilon}]\chi_{b}^{\epsilon}(x) + \ldots
\end{eqnarray}
where the fields $C_{abc}(x)$, $h_{ab}(x)$, $\chi_{b}^{\alpha}$ satisfy the linearized $D=11$ supergravity equations of motion and gauge invariances
\begin{eqnarray}
\partial^{c}[\partial_{c}h_{ab} - 2\partial_{(a}h_{b)c}] - \partial_{a}\partial_{b}h^{c}{}_{c} = 0\hspace{2mm}&,&\hspace{2mm} \delta h_{ab} = \partial_{(b}\Lambda_{c)}\nonumber\\
\partial^{d}\partial_{[a}C_{bcd]} = 0\hspace{2mm}&,&\hspace{2mm} \delta C_{abc} = \partial_{[a}\Lambda_{bc]}\nonumber\\
(\gamma^{abc})_{\alpha\beta}\partial_{b}\chi_{c}^{\beta} = 0 \hspace{2mm}&,&\hspace{2mm} \delta \chi_{a}^{\beta} = \partial_{a}\Lambda^{\beta}
\end{eqnarray}
where $\Lambda_{a}$, $\Lambda_{bc}$, $\Lambda^{\alpha}$ are arbitrary gauge parameters. 

The other BV fields of linearized $D=11$ supergravity are placed into different ghost sectors up to ghost number 7. 

Following 
$D=10$ dimensions \cite{Bjornsson:2010wm, Bjornsson:2010wu}, one can attempt to 
define a pure spinor measure from the eleven-dimensional scalar top cohomology, namely $\langle \lambda^{7}\theta^{9} \rangle = 1$ in order to give a consistent prescription for computing $N$-point correlation functions. This measure is easily shown to be BRST-invariant and supersymmetric and has already been successfully used to get the kinetic terms of the $D=11$ supergravity action from a second-quantized point of view \cite{Berkovits:2002uc}. Using this one can then propose that the N-point amplitude should be given by a correlation function of the form  \cite{Anguelova:2004pg}
\begin{eqnarray}\label{eq2}
\mathcal{A}^{11D}_{N} &=& \langle U^{(3)}_{1}(\tau_{1})U^{(3)}_{2}(\tau_{2})U^{(1)}_{3}(\tau_{3})\int d\tau_{4}V^{(0)}_{4}(\tau_{4})\ldots \int d\tau_{N} V^{(0)}_{N}(\tau_{N})\rangle
\end{eqnarray}
In this expression $U^{(3)}$ is the ghost number three vertex operator described above, $U^{(1)}$ is a ghost number one vertex operator and $V^{(0)}$ is a vertex operator of ghost number zero. Although it is possible to write an alternative prescription involving the ghost number four vertex operator containing the antifields of the $D=11$ supergravity physical fields, the existence of $V^{(0)}$ clearly plays a crucial role for the computation of the N-point correlation functions beyond $N=3$ in this framework. 

Having established the importance of the ghost number one and zero vertex operators, we now discuss their construction.


\section{Ghost number one vertex operator}\label{sec3}
The ghost number one vertex operator will be constructed from a small perturbation of the pure spinor BRST operator whose nilpotency will follow from the $D=11$ linearized supergravity equations of motion and the pure spinor constraint. We give a detailed review of the superspace formulation of $D=11$ supergravity in Appendix \ref{apA}\footnote{The linearized description of it can be readily obtained by dropping out interacting terms in the equations of motion displayed in this Appendix.}. Let us write the eleven-dimensional vielbeins in their linearized form
\begin{eqnarray}
E^{A} = E_{0}^{A} + h^{A} = (Dx^{a} + h^{a}{}_{b}Dx^{b} + h^{a}{}_{\beta}d\theta^{\beta}, d\theta^{\alpha} + h^{\alpha}{}_{\beta}d\theta^{\beta} + \psi_{b}^{\alpha}Dx^{b})
\end{eqnarray}
where
\begin{eqnarray}
Dx^{b} = dx^{b} + \frac{1}{2}(\theta\Gamma^{b}d\theta) \hspace{2mm}&,&\hspace{2mm} D_{\alpha} = \partial_{\alpha} - \frac{1}{2}(\Gamma^{c}\theta)_{\alpha}\partial_{c}
\end{eqnarray}
These give dually to first order
\begin{eqnarray}\label{0002}
\mathcal{D}_{\alpha} = D_{\alpha} - h_{\alpha}{}^{\beta}D_{\beta} - h_{\alpha}{}^{a}\partial_{a} \hspace{2mm}&,&\hspace{2mm} \mathcal{D}_{a} = \partial_{a} - \psi_{a}^{\alpha}D_{\alpha} - h_{a}{}^{b}\partial_{b}
\end{eqnarray}
On the other hand, using eqn. \eqref{eq12} one can show that at linear order
\begin{eqnarray}
[\mathcal{D}_{C}, \mathcal{D}_{D}\} &=& T_{CD}{}^{A}\mathcal{D}_{A} - 2\Omega_{[CD\}}{}^{A}\mathcal{D}_{A}\label{eq5}
\end{eqnarray}
where $[\cdot,\cdot\}$ is the graded (anti)commutator. Using the $D=11$ supergravity constraints \eqref{eq18}, one then finds that
\begin{eqnarray}\label{eq3}
\{\mathcal{D}_{\alpha}, \mathcal{D}_{\beta}\} &=& (\Gamma^{a})_{\alpha\beta}\mathcal{D}_{a} - 2\Omega_{(\alpha\beta)}^{\hspace{5mm}\gamma}\mathcal{D}_{\gamma}
\end{eqnarray}
Thus if one defines the BRST operator to be
\begin{eqnarray}\label{eq4}
Q &=& \lambda^{\alpha}(\mathcal{D}_{\alpha} + \Omega_{\alpha\beta}^{\hspace{4mm}\gamma}\lambda^{\beta}\frac{\partial}{\partial\lambda_{\gamma}})
\end{eqnarray}
then its nilpotency property immediately follows from the e.o.m \eqref{eq22}
\begin{eqnarray}\label{eq8}
\{Q, Q\} = \lambda^{\alpha}\lambda^{\beta}\lambda^{\delta}R_{(\alpha\beta\delta)}{}^{\epsilon}\frac{\partial}{\partial\lambda^{\epsilon}} = 0
\end{eqnarray} 

After converting \eqref{eq4} into a worldline vector with ghost number 1 by replacing operators by corresponding worldline fields, one concludes that
\begin{eqnarray}
Q = Q_{0} + U^{(1)} + \ldots \hspace{2mm}&,&\hspace{2mm} Q_{0} = \lambda^{\alpha}d_{\alpha}
\end{eqnarray}
where
\begin{eqnarray}\label{eq60}
U^{(1)} &=& \lambda^{\alpha}(h_{\alpha}{}^{a}P_{a} - h_{\alpha}{}^{\beta}d_{\beta} + \Omega_{\alpha\beta}^{\hspace{4mm}\gamma}N_{\gamma}{}^{\beta})
\end{eqnarray}
and $\dots$ means higher order terms. Thus $\{Q, Q\} = 0$  yields directly
\begin{eqnarray}
\{Q_{0} , U^{(1)}\} &=& 0
\end{eqnarray}
as desired.

The e.o.m satisfied by the superfields in \eqref{eq60} can be easily found by plugging \eqref{0002} into \eqref{eq5}. From the relation $\{\mathcal{D}_{\alpha}, \mathcal{D}_{\beta}\}$, one gets
\begin{eqnarray}
2 D_{(\alpha}h_{\beta)}{}^{a} + 2 h_{(\alpha}{}^{\delta}(\Gamma^{a})_{\beta)\delta} - h_{b}{}^{a}(\Gamma^{b})_{\alpha\beta} &=& 0\label{eq70}\\
2D_{(\alpha}h_{\beta)}{}^{\delta} - 2\Omega_{(\alpha\beta)}{}^{\delta} - (\Gamma^{a})_{\alpha\beta}\psi_{a}{}^{\delta} &=& 0\label{eq71}
\end{eqnarray}
From the relation $\{\mathcal{D}_{a}, \mathcal{D}_{\alpha}\}$ one finds
\begin{eqnarray}
\partial_{a}h_{\alpha}{}^{\beta} - D_{\alpha}\psi_{a}{}^{\beta} + T_{a\alpha}{}^{\beta} - 2\Omega_{a\alpha}{}^{\beta} &=& 0\label{eq82}\\
\partial_{a}h_{\alpha}{}^{b} - D_{\alpha}h_{a}{}^{b} + h_{a}{}^{\beta}(\Gamma^{b})_{\beta\alpha} &=& 0\label{eq80}
\end{eqnarray}
From the relation $\{\mathcal{D}_{a}, \mathcal{D}_{b}\}$ one obtains
\begin{eqnarray}
\partial_{a}\psi_{b}{}^{\alpha} - \partial_{b}\psi_{a}{}^{\alpha} + T_{ab}{}^{\alpha} &=& 0\\
\partial_{a}h_{b}{}^{c} - \partial_{b}h_{a}{}^{c} - 2\Omega_{[ab]}{}^{c} &=& 0
\end{eqnarray}
Moreover, the linearized supercurvature components can be written in terms of the super spin-conection using eqn. \eqref{eq11}
\begin{eqnarray}
R_{\alpha\beta c}{}^{d} &=& 2 D_{(\alpha}\Omega_{\beta)c}{}^{d}\label{eq72}\\
R_{a\alpha b}{}^{c} &=& \partial_{a}\Omega_{\alpha b}{}^{c} - D_{\alpha}\Omega_{ab}{}^{c}\\
R_{abc}{}^{d} &=& 2\partial_{[a}\Omega_{b]c}{}^{d}
\end{eqnarray}
As a consistency check, one can verify that $\{Q_{0}, U^{(1)}\} = 0$ as a consequence of the e.o.m \eqref{eq70}, \eqref{eq71}, \eqref{eq72}.

\section{Ghost number zero vertex operator}\label{sec4}
In order for a consistent standard equation to be satisfied, a ghost number zero vertex operator should exist and satisfy the relation
\begin{eqnarray}\label{eq75}
\{Q_{0}, V^{(0)}\} &=& P^{a}\partial_{a}U^{(1)}
\end{eqnarray}
where $U^{(1)}$ is the ghost number one vertex operator discussed above. To solve eqn. \eqref{eq75}, let us write first the most general ghost number zero vertex operator which is gauge invariant under the pure spinor constraint
\begin{eqnarray}\label{eq76}
V^{(0)} &=& P^{a}P^{b}\mathcal{G}_{ab} + P^{a}d_{\beta}\Psi_{a}^{\beta} + P^{a}N^{bc}\mathcal{W}_{abc} + d_{\alpha}d_{\beta}\mathcal{P}^{\alpha\beta} + d_{\alpha}N^{ab}\mathcal{T}_{ab}{}^{\alpha} + N^{ab}N^{cd}\mathcal{R}_{ab,cd}\nonumber\\
\end{eqnarray}
One can now compute the e.o.m that the superfields in \eqref{eq76} should satisfy such that \eqref{eq75} holds. After some algebraic manipulations one finds that
\begin{eqnarray}
\lambda^{\alpha}P^{a}P^{b}[D_{\alpha}\mathcal{G}_{ab} - \Psi_{(a}^{\beta}(\Gamma_{b)})_{\alpha\beta} - \partial_{a}h_{\alpha b}] &=& 0\label{eq100}\\
\lambda^{\alpha}P^{a}d_{\beta}[-D_{\alpha}\Psi_{a}^{\beta} - \frac{1}{2}\mathcal{W}_{a}{}^{bc}(\Gamma_{bc})^{\beta}{}_{\alpha} - 2(\Gamma_{a})_{\alpha\gamma}\mathcal{P}^{\gamma\beta} + \partial_{a}h_{\alpha}{}^{\beta}] &=& 0\label{eq81}\\
\lambda^{\alpha}P^{a}N^{bc}[D_{\alpha}\mathcal{W}_{abc} - (\Gamma_{a})_{\alpha\beta}\mathcal{T}_{bc}{}^{\beta} - \partial_{a}\Omega_{\alpha bc}] &=& 0\label{004}\\
\lambda^{\alpha}d_{\beta}d_{\gamma}[D_{\alpha}\mathcal{P}^{\beta\gamma} + \frac{1}{2}(\Gamma^{ab})^{\gamma}_{\hspace{2mm}\alpha}\mathcal{T}_{ab}{}^{\beta}] &=& 0 \label{001}\\
\lambda^{\beta}d_{\alpha}N^{ab}[D_{\beta}\mathcal{T}_{ab}{}^{\alpha} + \frac{1}{2}(\Gamma^{cd})^{\alpha}_{\hspace{2mm}\beta}\mathcal{R}_{cdab} + \frac{1}{2}(\Gamma^{cd})^{\alpha}{}_{\beta}\mathcal{R}_{abcd}] &=& 0\label{003}\\
\lambda^{\alpha}N^{ab}N^{cd}D_{\alpha}\mathcal{R}_{abcd} &=& 0\label{007}
\end{eqnarray}
The first equation can be automatically solved if one identifies $\mathcal{G}_{ab} = h_{ab}$, $\Psi_{a}^{\alpha} = \psi_{a}^{\alpha}$ as can be seen from \eqref{eq80}. Replacing this into \eqref{eq81} one gets
\begin{eqnarray}
\lambda^{\alpha}P^{a}d_{\beta}[-D_{\alpha}\psi_{a}^{\beta} - \frac{1}{2}\mathcal{W}_{a}{}^{bc}(\Gamma_{bc})^{\beta}{}_{\alpha} - 2(\Gamma_{a})_{\alpha\gamma}\mathcal{P}^{\gamma\beta} + \partial_{a}h_{\alpha}{}^{\beta}] &=& 0
\end{eqnarray}
After taking a look at eqn. \eqref{eq82}, one concludes that this equation becomes an identity if one identifies $\mathcal{W}_{abc} = \Omega_{abc}$, $-2(\Gamma_{a})_{\alpha\gamma}\mathcal{P}^{\gamma\beta} = T_{\alpha a}{}^{\beta}$. However this solution for $\mathcal{P}^{\alpha\beta}$ is inconsistent as will be shown now. If this identification were true, it would imply that
\begin{eqnarray}\label{eq85}
\mathcal{P}^{\alpha\beta} &=& -\frac{5}{192}(\Gamma^{abcd})^{\alpha\beta}H_{abcd}
\end{eqnarray}
If one now tries to recover $T_{\delta a}{}^{\beta}$ by multiplying eqn. \eqref{eq85} by $-2(\Gamma_{a})_{\delta\alpha}$, one finds that
\begin{eqnarray}
-2(\Gamma_{a})_{\delta\alpha}P^{\alpha\beta} 
&=& -\frac{5}{24}[(\Gamma^{cde})_{\delta}{}^{\beta}H_{acde} + \frac{1}{4}(\Gamma^{abcde})_{\delta}{}^{\beta}H_{bcde}]
\end{eqnarray}
which is clearly an inconsistency because of the eleven-dimensional structure of maximal supergravity (see eqn. \eqref{eq24}). 

Further evidence that $D=11$ supergravity is inconsistent with eqns. \eqref{eq100}-\eqref{007} can be found when trying to solve eqn. \eqref{001}. To see this, let us identify $\mathcal{T}_{ab}{}^{\alpha}$ with one the $D=11$ supergravity fields. Using dimensional analysis arguments one concludes that the most general expression for $\mathcal{T}_{ab}{}^{\alpha}$ should have the form
\begin{eqnarray}\label{eq91}
\mathcal{T}_{ab}{}^{\alpha} &=& T_{ab}{}^{\alpha} + a_{1}(\Gamma_{[a|c|})^{\alpha}{}_{\delta}T_{b]c}{}^{\delta} + a_{2}(\Gamma_{abcd})^{\alpha}{}_{\delta}T_{cd}{}^{\delta}
\end{eqnarray}
where $a_{1}$, $a_{2}$ are numerical constants to be determined. Using eqn. \eqref{eq90}, one can relate the last two terms on the right hand side of \eqref{eq91} to the first one, since $(\Gamma_{[a|c|})^{\alpha}{}_{\delta}T_{b]c}{}^{\delta} = T_{ab}{}^{\alpha}$ and $(\Gamma_{abcd})^{\alpha}{}_{\delta}T_{cd}{}^{\delta} = 2T_{ab}{}^{\alpha}$. This implies that $\mathcal{T}_{ab}{}^{\alpha} = b_{1}T_{ab}{}^{\alpha}$ where $b_{1}$ is a constant normalization factor. After plugging this and eqn. \eqref{eq85} into \eqref{001} one demonstrates that
\begin{eqnarray}
\lambda^{\alpha}d_{\beta}d_{\gamma}[-\frac{5}{96}(\Gamma^{bcde})^{\beta\gamma}D_{\alpha}H_{bcde} + b_{1}(\Gamma^{ab})^{\gamma}{}_{\alpha}T_{ab}{}^{\beta}] &=& 0
\end{eqnarray}
Since this equation is antisymmetric in $(\beta,\gamma)$, it should be true for the all antisymmetric gamma matrix projections of it, namely $C_{\beta\gamma}$, $(\Gamma^{fgh})_{\beta\gamma}$, $(\Gamma^{fghi})_{\beta\gamma}$. In particular, the 3-form projection requires 
\begin{eqnarray}
(\Gamma_{fgh})_{\beta\gamma}(\Gamma^{ab})_{\alpha}{}^{\gamma}T_{ab}{}^{\beta} &=& 0
\end{eqnarray}
However, the use of eqn. \eqref{eq90} allows one to show that 
\begin{eqnarray}
(\Gamma_{fgh})_{\beta\gamma}(\Gamma^{ab})_{\alpha}{}^{\gamma}T_{ab}{}^{\beta} &=& 24(\Gamma_{[h})_{\alpha\beta}T_{fg]}{}^{\beta}
\end{eqnarray}
which is non-zero and thus inconsistent with \eqref{001}. Thus  it is not possible to obtain a ghost number zero vertex operator from the $D=11$ supergravity fields that satisfy the standard descent equation \eqref{eq75}.

\section{Discussion}
In this paper we have constructed a ghost number one vertex operator involving more terms in its definition compared to that presented in \cite{Anguelova:2004pg}. In principle, there is no physical reason to ignore them in the 3-point function computations, so it would be interesting to see how they affect the results found in \cite{Anguelova:2004pg}. Moreover, an explicit relation between $U^{(1)}$ and $U^{(3)}$ would be important to understand the structures underlying the eleven-dimensional pure spinor framework, for example to prove permutation invariance of the correlator. A first step in this direction was provided in \citep{Berkovits:2018gbq} where it was shown that
\begin{eqnarray}
\partial_{a}U^{(3)} &=& (\lambda\Gamma_{ab}\lambda)\Phi^{b} + Q(\Xi_{a})
\end{eqnarray}
where $\Phi^{b} = \lambda^{\alpha}h_{\alpha}{}^{b}$ and $\Xi_{a}$ is a ghost number two operator. It should be emphasized that this superfield $\Phi^{b}$ has been successfully used to study $D=11$ supergravity from a second-quantized perspective \cite{Cederwall:2010tn,Cederwall:2012es,Berkovits:2018gbq}. In particular, note that $\Phi^{b}$ is contained in $U^{(1)}$ after contraction with the momentum $P_{b}$. 

On the other hand, we have shown that it is not possible to write a ghost number zero vertex operator made out of the $D=11$ supergravity superfields satisfying a standard descent equation. One possible resolution is to extend the present framework to its non-minimal version by introducing the standard non-minimal pure spinor variables. In this setting, a ghost number zero vertex operator can be defined using the relation
\begin{eqnarray}\label{dis1}
\{b, U^{(1)}\} &=& \tilde{V}^{(0)}
\end{eqnarray} 
where $b$ was found in \cite{Cederwall:2012es} and simplified in \cite{Berkovits:2017xst} where it was shown that $b$ is nilpotent up to BRST-exact terms. Since $b$ satisfies $\{Q, b\} = H$, $\tilde{V}^{(0)}$ satisfies the standard descent equation
\begin{eqnarray}
[Q, \tilde{V}^{(0)}] &=& [H, U^{(1)}]
\end{eqnarray}
Unlike the ten-dimensional case \cite{Bjornsson:2010wu}, $\tilde{V}^{(0)}$ cannot be split into a function depending only on minimal variables plus a BRST-exact term as follows from our result in section \ref{sec4}. This implies that $\tilde{V}^{(0)}$ will be a complicated function depending on minimal and non-minimal variables. It would be interesting to determine whether  the non-minimal sector decouples from the theory perhaps with the use of some appropriate measure in the correlation functions. 

Another approach is to impose additional constraints on the eleven-dimensional pure spinor in such a way that more terms in the ghost number zero vertex operator are allowed. For example, one could impose the full Cartan purity condition $\lambda\Gamma^{ab}\lambda=0$ as was considered in \cite{Howe:1991bx,Babalic:2008ga}, and it would be interesting to see if there is some relation of this constraint with the present work.

\section{Acknowledgments}
NB acknowledges FAPESP grants 2016/01343-7 and 2014/18634-9 and CNPq grant 300256/94-9 for partial financial support. MG acknowledges FAPESP grants 15/23732-2 and 18/10159-0 for financial support. EC and LJM acknowledge support from the EPSRC grant EP/M018911/1. MG would also like to thank the hospitality of Mathematical Institute at Oxford where final stages of this research were done and Perimeter Institute for their hospitality during various stages of preparation of the manuscript. This research is supported in part by U.S. Department of Energy grant DE-SC0009999 and by funds provided by the University of California.

\begin{appendices}
\section{Review of superspace formulation of $D=11$ supergravity}\label{apA}
In this Appendix we will review the original superspace formulation of $D=11$ supergravity given in \cite{Brink:1980az}. This turns out to be useful to fix conventions and get consistently the equations of motion for the dynamical superfields, which in turn play a crucial role when constructing the ghost number one and zero vertex operators of sections \ref{sec3} and \ref{sec4}.  Let us start by fixing notation. Latin capital letters from the beginning/middle of the alphabet will be used to represent tangent/coordinate superspace indices. The vielbein and spin-connection will be defined to be 1-forms on superspace as follows
\begin{eqnarray}
E^{A} = d Z^{M}E_{M}{}^{A} \hspace{2mm}&,&\hspace{2mm} \Omega_{A}{}^{B} = dZ^{M}\Omega_{MA}{}^{B}
\end{eqnarray}
where $d Z^{M} = (dX^{m}, d\theta^{\mu})$. The existence of $\Omega_{A}{}^{B}$ allows one to introduce a super covariant derivative which will act on an arbitrary tensor $\mathcal{F}_{A_{1}\ldots A_{m}}{}^{B_{1}\ldots B_{n}}$ in the form
\begin{eqnarray}\label{eq12}
\mathcal{D}\mathcal{F}_{A_{1}\ldots A_{m}}{}^{B_{1}\ldots B_{n}} &=& d \mathcal{F}_{A_{1}\ldots A_{m}}{}^{B_{1}\ldots B_{n}} - \Omega_{A_{1}}{}^{C}\mathcal{F}_{C\ldots A_{m}}{}^{B_{1}\ldots B_{n}} - \ldots + \mathcal{F}_{A_{1}\ldots A_{m}}{}^{C\ldots B_{n}}\Omega_{C}{}^{B_{1}} + \ldots\nonumber\\
\end{eqnarray}
where $d$ is the standard exterior derivative. Next one introduces the 2-form supertorsion as the covariant derivative of the 1-form supervielbein
\begin{eqnarray}\label{eq9}
T^{A} &=& \frac{1}{2}E^{A}E^{B}T_{BA}{}^{C} = \mathcal{D}E^{A}\nonumber\\
&=& d E^{A} + E^{B}\Omega_{B}{}^{A}
\end{eqnarray}
and the 2-form supercurvature as the covariant derivative of the 1-form super spin-connection
\begin{eqnarray}\label{eq11}
R_{A}{}^{B} &=& \frac{1}{2}E^{C}E^{D}R_{DC,A}{}^{B} = \mathcal{D}\Omega_{A}{}^{B}\nonumber\\
&=& d\Omega_{A}{}^{B} + \Omega_{A}{}^{C}\Omega_{C}{}^{B}
\end{eqnarray}
As usual, we will constrain the super spin-connection components to satisfy
\begin{eqnarray}
\Omega_{\alpha\beta} &=& \frac{1}{4}(\Gamma^{mn})_{\alpha\beta}\Omega_{mn}
\end{eqnarray}
and all the other components to vanish. This choice automatically implies that
\begin{eqnarray}\label{eq21}
R_{DC,\alpha\beta} &=& \frac{1}{4}(\Gamma^{mn})_{\alpha\beta}R_{DC,mn}
\end{eqnarray}
Using \eqref{eq12}, \eqref{eq9}, \eqref{eq11} one easily finds the so-called Bianchi identities
\begin{eqnarray}
\mathcal{D}T^{A} = E^{B} R_{B}{}^{A} \hspace{2mm}&,&\hspace{2mm} \mathcal{D}R_{A}{}^{B} = 0
\end{eqnarray}
which in component notation read
\begin{eqnarray}
R_{[BD,C\}}{}^{A} - \nabla_{[B}T_{DC\}}{}^{A} - T_{[BD}{}^{F}T_{|F|C\}}{}^{A} &=& 0\label{eq15}\\
\nabla_{[F}R_{DC\},A}{}^{B} + T_{[FD}{}^{E}R_{|E|C\},A}{}^{B} &=& 0\label{eq16}
\end{eqnarray}
where $[\cdot,\cdot\}$ is a graded antisymmetrization. 

\vspace{2mm}
Furthermore, a 4-form superfield can be also introduced
\begin{eqnarray}
H &=& \frac{1}{4!}E^{D}E^{C}E^{B}E^{A}H_{ABCD}
\end{eqnarray}
which will be required to satisfy $dH = 0$. This condition gives rise to a new identity, which in component notation takes the form
\begin{eqnarray}\label{eq17}
\nabla_{[F}H_{ABCD\}} + 2T_{[FA}{}^{E}H_{|E|BCD\}} &=& 0
\end{eqnarray}

\vspace{2mm}
In order to put the theory on-shell we will impose the standard conventional and dynamical constraints, namely
\begin{eqnarray}\label{eq18}
H_{\alpha abc} = H_{\alpha\beta\delta a} = H_{\alpha\beta\delta\epsilon} = T_{ab}{}^{c} = T_{\alpha\beta}{}^{\delta} = T_{a\alpha}{}^{c} = 0\nonumber\\
T_{\alpha\beta}{}^{a} = (\Gamma^{a})_{\alpha\beta} \hspace{6mm},\hspace{6mm} H_{\alpha\beta ab} = (\Gamma_{ab})_{\alpha\beta}\hspace{10mm}
\end{eqnarray}
In this way, the only dynamical superfields of $D=11$ supergravity are $H_{abcd}$, $T_{a\alpha}{}^{\beta}$, $T_{ab}{}^{\alpha}$. To see how this works one should solve the identities \eqref{eq15}, \eqref{eq16}, \eqref{eq17} by plugging \eqref{eq18} into them. For instance, from eqn. \eqref{eq17} one gets
\begin{eqnarray}
(\alpha\beta\delta\gamma a)&:& \hspace{4mm} 3T_{(\alpha\beta}{}^{A}H_{|A|\delta\gamma)a} = 0\nonumber\\
&&\rightarrow 3 (\Gamma^{a})_{(\alpha\beta}(\Gamma_{ab})_{\delta\gamma)} = 0\\
(\alpha\beta cde)&:& \hspace{4mm}2[T_{\alpha\beta}{}^{A}H_{Acde} - 6T_{(\alpha [c}{}^{A}H_{|A|\beta) de]} + 3T_{[cd}{}^{A}H_{|A\alpha\beta| e]}] = 0\nonumber\\
&&\rightarrow (\Gamma^{a})_{\alpha\beta}H_{acde} - 6T_{(\alpha[c}{}^{\delta}(\Gamma_{de]})_{\beta)\delta} = 0\label{eq19}\\
(\alpha bcde) &:& \hspace{4mm}\nabla_{\alpha}H_{bcde} + 2(3 T_{[bc}{}^{\beta}H_{|\beta\alpha| de]}) = 0\nonumber\\
&& \rightarrow  \nabla_{\alpha}H_{bcde} + 6(\Gamma_{[de})_{\alpha\beta}T_{bc]}{}^{\beta} = 0\label{eq20}\\
(abcde)&:& \hspace{4mm}\nabla_{[a}H_{bcde]} = 0
\end{eqnarray}
The first equation is just a consistency check. The second equation \eqref{eq19} tells us that
\begin{eqnarray}
H_{abcd} &=& \frac{3}{16}(\Gamma^{a}\Gamma_{[de})^{\alpha}{}_{\delta}T_{|\alpha| c]}{}^{\delta}\nonumber\\
&=& \frac{3}{8}\eta_{a[d}(\Gamma_{e})^{\alpha}{}_{\delta}T_{|\alpha| c]}{}^{\delta} + \frac{3}{16}(\Gamma_{a[de})^{\alpha}{}_{\delta}T_{|\alpha|c]}{}^{\delta}
\end{eqnarray} 
which implies that $(\Gamma_{a})^{\alpha}{}_{\delta}T_{\alpha b}^{\delta} = 0$ and
\begin{eqnarray}
H_{abcd} &=& \frac{3}{16}(\Gamma_{a[de})^{\alpha}{}_{\delta}T_{|\alpha|c]}{}^{\delta}
\end{eqnarray}
This implies that $T_{\alpha a}{}^{\beta}$ can be written in terms of $H_{abcd}$. Using symmetry arguments one finds that
\begin{eqnarray}
T_{\alpha a}{}^{\delta} &=& c_{3}(\Gamma^{bcd})_{\alpha}{}^{\delta}H_{abcd} + c_{5}(\Gamma_{abcde})_{\alpha}{}^{\delta}H^{bcde}
\end{eqnarray}
The use of eqn. \eqref{eq19} tells us that
\begin{eqnarray}
\frac{1}{6}(\Gamma^{a})_{\alpha\beta}H_{acde} &=& c_{3}(\Gamma^{bfg}\Gamma_{[de})_{(\alpha\beta)}H_{c]bfg} + c_{5}(\Gamma_{[c|abfg|}\Gamma_{de]})_{(\alpha\beta)}H^{abfg}\nonumber\\
&=& -6c_{3}(\Gamma^{g})_{\alpha\beta}H_{cdeg} + c_{3}(\Gamma^{[de|bfg|})_{\alpha\beta}H_{c]bfg} - 8c_{5}(\Gamma^{[de|bfg|})_{\alpha\beta}H_{c]bfg}\nonumber\\
\end{eqnarray}
which leads us to conclude that $c_{3} = \frac{1}{36}$ and $c_{5} = \frac{c_{3}}{8} = \frac{1}{288}$. Thus one can write
\begin{eqnarray}\label{eq24}
T_{\alpha a}{}^{\delta} &=& \frac{1}{36}[(\Gamma^{bcd})_{\alpha}{}^{\delta}H_{abcd} + \frac{1}{8}(\Gamma_{abcde})_{\alpha}{}^{\delta}H^{bcde}]
\end{eqnarray}
Moreover, $H_{abcd}$ and $T_{ab}{}^{\alpha}$ are related to each other via eqn. \eqref{eq20}
\begin{eqnarray}
\nabla_{\alpha}H_{bcde} &=& -6(\Gamma_{[de})_{\alpha\beta}T_{bc]}{}^{\beta}
\end{eqnarray}
Next, one can use the Bianchi identity \eqref{eq15} together with eqn. \eqref{eq21} to find
\begin{eqnarray}
(\alpha\beta\delta)(\gamma)&:& \hspace{4mm} \frac{1}{4}(\Gamma^{ab})_{(\delta}{}^{\gamma}R_{\alpha\beta),ab} + (\Gamma^{a})_{(\alpha\beta}T_{\delta)a}{}^{\gamma} = 0\label{eq22}\\
(a\alpha\beta)(\gamma)&:&\hspace{4mm}(\Gamma^{bc})_{(\beta}{}^{\gamma}R_{|a|\alpha),bc} - 4\nabla_{(\alpha}T_{\beta)a}{}^{\gamma} - 2(\Gamma^{b})_{\alpha\beta}T_{ba}{}^{\gamma} = 0\label{eq26}\\
(\alpha\beta b)(c) &:& \hspace{4mm} R_{(\alpha\beta),b}{}^{c} + 2(\Gamma^{c})_{\gamma(\beta}T_{\alpha)b}{}^{\gamma} = 0\label{eq23}\\
(ab\alpha)(\beta)&:& \hspace{4mm}\frac{1}{4}(\Gamma^{cd})_{\alpha}{}^{\beta}R_{ab,cd} + 2\nabla_{[a}T_{|\alpha|b]}{}^{\beta} - \nabla_{\alpha}T_{ab}{}^{\beta} - 2T_{\alpha [a}{}^{\delta}T_{|\delta| b]}{}^{\beta} = 0\label{eq40}\\
(\alpha ab)(c)&:& \hspace{4mm} R_{\alpha [a,b]}{}^{c} - \frac{1}{2}(\Gamma^{c})_{\gamma\alpha}T_{ab}{}^{\gamma} = 0\label{eq27}\\
(abc)(\alpha)&:&\hspace{4mm} \nabla_{[a}T_{bc]}{}^{\alpha} + T_{[ab}{}^{\gamma}T_{|\gamma|c]}{}^{\alpha} = 0\\
(abc)(d)&:& \hspace{4mm} R_{[ab,c]}{}^{d} = 0
\end{eqnarray}
The eqns. \eqref{eq22}, \eqref{eq23} imply that 
\begin{eqnarray}\label{eq25}
\frac{1}{2}(\Gamma^{ab})_{(\delta}{}^{\gamma}(\Gamma_{a})_{|\epsilon|\beta}T_{\alpha)b}{}^{\epsilon} + (\Gamma^{a})_{(\alpha\beta}T_{\delta)a}{}^{\gamma} &=& 0
\end{eqnarray}
After replacing \eqref{eq24} in \eqref{eq25}, one gets
\begin{eqnarray}\label{eq255}
H_{cdef}[\frac{3}{2}(\Gamma^{cd})_{(\delta}{}^{\gamma}(\Gamma^{ef})_{\alpha\beta)} + \frac{1}{16}(\Gamma^{ab})_{(\delta}{}^{\gamma}(\Gamma_{abcdef})_{\alpha\beta)} - (\Gamma^{c})_{(\alpha\beta}(\Gamma^{def})_{\delta)}{}^{\gamma} - \frac{1}{8}(\Gamma^{a})_{(\alpha\beta}(\Gamma_{acdef})_{\delta)}{}^{\gamma}] = 0\nonumber\\
\end{eqnarray}
which is an identity as can be shown by multiplying on both sides of \eqref{eq255} by $(\Gamma^{a})^{\alpha}$, $(\Gamma^{ab})^{\alpha\beta}$, $(\Gamma^{abcde})^{\alpha\beta}$\footnote{The GAMMA package \cite{Gran:2001yh} turns out to be useful for this type of computations.}.

\vspace{2mm}
Using eqns. \eqref{eq26}, \eqref{eq27} one gets a set of constraints on $T_{ab}{}^{\alpha}$. Let us see how this works. The use of eqn. \eqref{eq27} allows us to write
\begin{eqnarray}
(\Gamma^{bc})_{\beta}{}^{\gamma}R_{a\alpha, bc} &=& - (\Gamma^{bc})_{\beta}{}^{\gamma}(\Gamma_{c})_{\alpha\delta}T_{ab}{}^{\delta} + \frac{1}{2}(\Gamma^{bc})_{\beta}{}^{\gamma}(\Gamma_{a})_{\alpha\delta}T_{bc}{}^{\delta}
\end{eqnarray}
Plugging this expression into eqn. \eqref{eq26} ones arrives at the relation
\begin{eqnarray}\label{eq256}
- (\Gamma^{bc})_{(\beta}{}^{\gamma}(\Gamma_{c})_{\alpha)\delta}T_{ab}{}^{\delta} + \frac{1}{2}(\Gamma^{bc})_{(\beta}{}^{\gamma}(\Gamma_{a})_{\alpha)\delta}T_{bc}{}^{\delta} - 4\nabla_{(\alpha}T_{\beta)a}{}^{\gamma} - 2(\Gamma^{b})_{\alpha\beta}T_{ba}{}^{\gamma} &=& 0
\end{eqnarray}
Moreover, from eqns. \eqref{eq24}, \eqref{eq20} one finds
\begin{eqnarray}
\nabla_{(\alpha}T_{\beta)a}{}^{\gamma} &=& -\frac{1}{6}[(\Gamma^{bcd})_{(\beta}{}^{\gamma}(\Gamma_{[cd})_{\alpha)\delta}T_{ab]}{}^{\delta} + \frac{1}{8}(\Gamma_{abcde})_{(\beta}{}^{\gamma}(\Gamma_{de})_{\alpha)\delta}T_{bc}{}^{\delta}]
\end{eqnarray}
Thus eqn. \eqref{eq256} becomes
\begin{eqnarray}
- (\Gamma^{bc})_{(\beta}{}^{\gamma}(\Gamma_{c})_{\alpha)\delta}T_{ab}{}^{\delta} + \frac{1}{2}(\Gamma^{bc})_{(\beta}{}^{\gamma}(\Gamma_{a})_{\alpha)\delta}T_{bc}{}^{\delta} - 2(\Gamma^{b})_{\alpha\beta}T_{ba}{}^{\gamma}\nonumber\\
+ \frac{2}{3}[(\Gamma^{bcd})_{(\beta}{}^{\gamma}(\Gamma_{[cd})_{\alpha)\delta}T_{ab]}{}^{\delta} + \frac{1}{8}(\Gamma_{abcde})_{(\beta}{}^{\gamma}(\Gamma_{de})_{\alpha)\delta}T_{bc}{}^{\delta}] = 0
\end{eqnarray}
After multiplying on both sides by $(\Gamma^{a})^{\alpha\beta}$, $(\Gamma^{ab})^{\alpha\beta}$, $(\Gamma^{abcde})^{\alpha\beta}$ one arrives at
\begin{eqnarray}\label{eq90}
(\Gamma^{abc})_{\alpha\beta}T_{bc}{}^{\beta} = (\Gamma^{ab})_{\alpha\beta}T_{ab}{}^{\beta} = (\Gamma^{b})_{\alpha\beta}T_{ab}{}^{\beta} = 0
\end{eqnarray}
Using this result and eqn. \eqref{eq20} one learns that
\begin{eqnarray}\label{eq258}
(\Gamma^{cd}\Gamma_{[ab})_{\alpha\beta}T_{cd]}{}^{\beta} = -7 T_{ab\,\alpha} \hspace{2mm}&,& T_{ab}{}^{\alpha} = \frac{1}{42}(\Gamma^{cd})^{\alpha\beta}\nabla_{\beta}H_{abcd}
\end{eqnarray}
Plugging this back into eqn. \eqref{eq20}, one finds that
\begin{eqnarray}
\nabla_{\alpha}H_{abcd} &=& \frac{1}{7}(\Gamma_{[cd}\Gamma^{ef})_{\alpha}{}^{\delta}\nabla_{\delta}H_{ab]ef}
\end{eqnarray}

\vspace{2mm}
On the other hand, after multiplying by
$(\Gamma^{cd})_{\beta}{}^{\alpha}$ and $\eta^{bd}$ on both sides of eqn. \eqref{eq40} one obtains
\begin{eqnarray}\label{eq42}
R_{ac} &=& \frac{1}{8}\nabla_{\alpha}T_{ac}{}^{\alpha} - \frac{1}{8}(\Gamma^{cb})_{\beta}{}^{\alpha}T_{\alpha[a}{}^{\delta}T_{|\delta|b]}{}^{\beta}
\end{eqnarray}
The first term vanishes as a consequence of eqns. \eqref{eq23}, \eqref{eq24}. The second term in \eqref{eq42} takes the simple form
\begin{eqnarray}
(\Gamma^{cb})_{\beta}{}^{\alpha}T_{\alpha[a}{}^{\delta}T_{|\delta|b]}{}^{\beta} &=& \frac{2}{3}H_{adef}H_{c}{}^{def} - \frac{1}{18}\eta_{ac}H_{defg}H^{defg}
\end{eqnarray}
Thus, the graviton e.o.m is given by
\begin{eqnarray}
R_{ac} &=& -\frac{1}{12}H_{adef}H_{c}{}^{def} + \frac{1}{144}\eta_{ac}H_{defg}H^{defg}
\end{eqnarray}
Finally, one can obtain the e.o.m for the 4-form field strength by multiplying on both sides of eqn. \eqref{eq40} by $(\Gamma^{c})_{\beta}{}^{\alpha}$ and using \eqref{eq258} to get
\begin{eqnarray}
\frac{1}{42}(\Gamma^{c})_{\beta}{}^{\alpha}(\Gamma^{de})^{\beta\delta}\nabla_{\alpha}\nabla_{\delta}H_{abde} - \frac{1}{1296}\epsilon_{abcdefghijk}H^{defg}H^{hijk} &=& 0
\end{eqnarray}
So after antisymmetrizing in $(a,b,c)$ one concludes that
\begin{eqnarray}
\nabla^{d}H_{dabc} + \frac{1}{1192}\epsilon_{abcdefghijk}H^{defg}H^{hijk} &=& 0
\end{eqnarray}
where the identity $(\Gamma^{b})^{\beta\alpha}\nabla_{\alpha}H_{bcde} = - \frac{3}{7}(\Gamma_{[c})^{\beta}{}_{\delta}(\Gamma^{fg})^{\delta\gamma}\nabla_{\gamma}H_{de]fg}$ was used.

\end{appendices}
\providecommand{\href}[2]{#2}\begingroup\raggedright\endgroup


\end{document}